\documentclass{jfm}
\usepackage{natbib}
\usepackage{epsfig}
\usepackage{amssymb}

\def\begineq{\begin{equation}}
\def\endeq{\end{equation}}
\def\be{\begin{equation}}
\def\ee{\end{equation}}

\title[  Turbulence Models Generator.  .]
{ Turbulence Models Generator.   }
\author{Victor Yakhot }
\affiliation{`Department of Aerospace and Mechanical Engineering, \\
Boston University, Boston, MA 02215}
\begin{document}
\maketitle
 
\pubyear{2007}
\volume{999}
\pagerange{1--10}
\date{?? and in revised form ??}
\setcounter{page}{1}

\maketitle

\begin{abstract}
\noindent  In this paper we explore a  possibility that all  transport turbulent models are contained in a coarse-grained kinetic equation. 
Building on  a recent work by H.Chen  et al (2004),  we 
 account for   fluctuations  of a single -point probability density in turbulence,  by introducing    a``two-level''   ( ${\bf c,v}$)-phase-space,    separating microscopic (${\bf c'\equiv  c_{micro}= c-v}$)  and hydrodynamic (${\bf v'=v-V}$)  modes.  
Unlike   traditional kinetic theories,  with hydrodynamic approximations derived in terms of  small deviations   from thermodynamic equilibrium,   the theory   developed in this work,   is  based on a far- from -equilibrium   isotropic and homogeneous turbulence as an  unperturbed state.
The expansion in dimensionless rate of strain  leads to a new class  of turbulent models, including  the well-known ${\cal K}-{\cal E}$,  Reynolds stress  and all possible   nonlinear models.  The  role of interaction of the   fluxes in physical space  with the energy flux across the scales, not present in standard modeling,  is demonstrated  on example of turbulent  channel flow.  To close the system,  neither  equation for turbulent kinetic energy nor information on  pressure-velocity correlations,   contained in the derived coarse-grained   kinetic equation,   are  needed. 
\end{abstract}

\vspace{0.3in}

\section{Introduction}

\noindent  In this paper we revisit an  old  and all-important problem of turbulent modeling. The problem,   
first formulated  in terms  of  images borrowed from kinetic theory 
 by Prandtl (1925),  was  later developed by  Kolmogorov (1942),  Launder and Spaulding  (1974) and many  of their followers.  The success   of these semi-empirical approaches can hardly be overestimated:  with development of powerful computers turbulent modeling became an important part of scientific and engineering design process.  Still, these models,  involving low -order time - derivatives and low powers of dimensionless rate of strain,  are  less effective in describing strongly sheared  and rapidly distorted flows where the dimensionless rate of strain is not too small.   To improve  performance,  various non-linear models,  pioneered by Speziale (1987),  accounting for the next  order in powers of  the rate of strain,  have been proposed.   Since the second-order  trancation of   a lacking-small -parameter  expansion is problematic,  these models  had a mixed success. (The role of the large-eddy simulations (LES) in description of complex flows will be discussed in Conclusions. )

\noindent  The attempts to systematically derive turbulence transport models from the Navier-Stokes equations  (NS),  based on renormalized  Wyld's perturbation expansions (RNG (Yakhot/ Smith et al (1992),  Rubinstein and Barton (1990)),  double expansion (Yakhot  et al (1992)),  DIA (Yoshizawa (1987)),  were relatively successful and the resulting equations  found their place in engineering.  It became clear soon  that  resummation of the series  was impractical  and  the  approach was restricted   to   the low- orders.

\noindent It has been shown  recently   by H. Chen et al (2004) that  the turbulence  models,   derived  by  application of the Chapman-Enskog expansion (CE) to a {\it model}  kinetic (Boltzmann-BGK) equation written for hydrodynamic modes,   are similar to those previously obtained from the Wyld  (1961) expansion,  applied directly to  the Navier-Stokes equations. The main result of Chen et al  (2004) can be stated  as follows:   the perturbation theory,   leading  from kinetic equation to the Navier-Stokes (NS)  equations,  generates the well-known    turbulence models if,  instead of  the relaxation time $\tau= \nu_{0}/\theta=const$  of kinetic theory,  one writes   $\tau_{T}\approx K/{\cal E}$ and  represents the  temperature $\theta$ as $\theta=K$. Here $K$ and ${\cal E}$ are the mean turbulent  kinetic energy  and dissipation rates, respectively.  Thus,  the formal superficial similarity  between the two  perturbation expansions has  been established.  This  result means that, in principle,  the models of an arbitrary non-linearity and complexity,  which we even cannot explicitly  write down,   are contained in a simple kinetic equation.  

\noindent  In this work,   we would like to reformulate the procedure developed in Chen (2004)  by taking into account  both microscopic and hydrodynamic modes.  The coarse-grained  kinetic equation for hydrodynamic modes  is then obtained by integrating out the fast microscopic variables. 
To achieve this goal,  we had   to substantially redefine the  expansion  procedure and,  as a result,  obtain a qualitatively new class of turbulence models.

\subsection {The Boltzmann -BGK  equation.} 

\noindent Kinetic theory for a low-density gas  is based on the Boltzmann equation

\begin{equation}
\frac{\partial f}{\partial t}+{\bf c\cdot \nabla}f={\cal C}(f)
\end{equation}

\noindent  for a single particle distribution function $f({\bf c},{\bf x},t)$,  where collision integral

 \begin{equation}
{\cal C}(f) = \int w' (f'f'_1 - f\; f_1)\; d\Gamma_1 d\Gamma' d\Gamma'_1
= \int |c_{rel} |(f'f'_1 - f\; f_1) d\sigma d^3 p_{1}\equiv i_{in}-i_{out}
\label{coll}
\end{equation} 

\noindent  If the system Hamiltonin is known, this equation can be derived directly from the Liuville theorem.  Here the element of the phase -space volume $d\Gamma=d{\bf x} d{\bf p}$ and  $|c_{rel}|=|{\bf c_{1}-c}|$ is the relative velocity of colliding particles.  If the intermolecular distance is $\lambda>>a$ where $a$ is the length-scale of intermolecular interaction,  the strait particle trajectories  between collisions are  assumed in deriving expression (1.2). 
The probability density functions in (1.2) are $f'=f'(r,{\bf c'},t)$ and $f'_{1}=f'(r,{\bf c'_{1}},t)$. Since in accord with the theory of elastic collisions, the velocities ${\bf c', \ c'_{1}}$  are expressed in terms of ${\bf c,\ c_{1}}$, the integration in (1.2) is carried out over the phase space $\Gamma_{1}$ only.   
The  relaxation -time approximation (RTA):

\begin{equation}
{\cal C}=-\frac{f-f^{eq}}{\tau}
\end{equation}

\noindent  where $f^{eq}$ is the distribution function in thermodynamic equilibrium and $\tau$ is a properly chosen relaxation time,  is an often used anztaz,  mentioned in Landau and Lifshitz (1981) as a "rough estimate of the ( Boltzmann) collision integral" (1.2).  Still, recent implementations  of this approximation in the so called "Lattice Boltzmann" numerical codes, led to a remarkable  success in simulating a wide variety of extremely complex fluid flows.  
 As of today,  the approach has been tested on basically all examples of classic laminar flows  (Benzi (1992), Chen (1998),  Succi (2001)).  
 The robustness, speed and  simplicity of the method,  made it  an attractive tool for both theoretical investigations and engineering design. 
 
\noindent Even more spectacular,  and somewhat  unexpected,  is  success enjoyed by the Lattice Boltzmann method   in simulating complex turbulent flows (H. Chen (2003)).  In this application,   the relaxation time $\tau$ is expressed in terms  of hydrodynamic observables:  $\tau\approx {\cal K}/{\cal E}$ , where turbulent kinetic energy is ${\cal K}=v_{rms}^{2}/2$ and the mean dissipation rate is ${\cal E}=\nu\overline{(\frac{\partial v_{i}}{\partial x_{j}})^{2}}$.   In this case,   in  the first  order of  the Chapman-Enskog  (CE)  expansion one obtains  the well-known ${\cal K}-{\cal E}$ model with the non-linear ${\cal K}-{\cal E}$ model appearing  in the next  order (Chen et al (2004)). 
In addition to  speed and simplicity,  the most attractive feature of this type of turbulence modeling is that it  {\it does not require theory  of  pressure-velocity correlations}  which,  in complex flows,  is an extremely difficult,  still unsolved,  problem.  

\noindent  
Typically,  derivation  of hydrodynamic Navier-Stokes  equations  from kinetic theory is based on a few assumptions. 1.~ If  spatial gradients of mean velocity,  density  and temperature  are equal to zero, the gas  is assumed in  thermodynamic equilibrium;  2.~ Expansion in powers 
of dimensionless rate of strain $\eta\equiv \eta_{ij}=\tau S_{ij}$ where $S_{ij}=\frac{1}{2}(\frac{\partial V_{i}}{\partial x_{j}}+\frac{\partial V_{j}}{\partial x_{i}})$  leads  to transport equations describing various physical phenomena.

\noindent  The equations,  derived   this way can be  used to fully describe   large Reynolds number turbulent flows  requiring   at least $N\approx Re^{3}$ number of degrees of freedom. In the flows of practical importance this number is huge and 
therefore, the value of a   coarse graining procedure leading 
 to  turbulence models  can hardly be overestimated . 

\noindent It has been shown (Yakhot et al (1992) ) that the systematic derivation of turbulence models directly  from the Navier-Stokes equations,  can be  formulated in terms of  Wyld's (1961)  diagramamtic  expansion  in  {\it two}  dimensionless parameters:  the Reynolds number  $Re=u_{rms}L/\nu_{0}$ and, familiar from kinetic theory,  dimensionless rate of strain $\eta=\tau_{T}S_{ij}$.  The first expansion  is responsible for description  of isotropic and homogeneous turbulence with the non-zero energy flux  across the scales and the second ~-~ for the non-zero {\it spatial}  fluxes and flow structures.  Due to proliferation of tensorial indices in the high-order contributions, the resummation of the   expansion  is an extremely difficult task and  it has been conjectured by Polyakov  (2001) in the middle of seventies that it may result in a kinetic equation containing all terms of Wyld's  series.   If this is true,  a  turbulence model of an arbitrary non-linearity and complexity may be contained in a relatively simple kinetic equation. {\it The most important and interesting feature of this system is that,  unlike equilibrium,   the $\eta=0$ state  of isotropic and homogeneous turbulence is not flux- free but involves a large $O(1)$ energy flux across the scales. Thus,  to derive turbulence models,  it is desirable to develop a kinetic approach not based on the equilibrium, flux-free, zero-order state. }

\noindent In this paper we,  building upon a remarkable work of H. Chen et  al (2004),   consider both turbulent (hydrodynamic) and thermal  (microscopic) velocity fluctuations  
as governed by the Boltzmann -BGK kinetic equation.  Since microscopic  and hydrodynamic fluctuations occupy  their respective fractions  of the phase-space,   this equation is formally  defined on a  somewhat enlarged  phase-space.  The coarse-graining, eliminates the small-scale fast microscopic fluctuations and restores  the six-dimensional space,  leading   to the kinetic equation for the {\it  large-scale (turbulent) velocity  component only}.  In this case,  the zero-order,   zero-mean-spatial-gradient state is not an equilibrium but that of isotropic and homogeneous turbulence characterized by a finite energy  flux across the scales.  The Chapman-Enskog expansion applied to this state leads  to a  novel set of turbulence models of arbitrary nonlinearity. 

\noindent This paper is organized as follows. In the next Section we, for the sake of clarity and continuity,   describe traditional derivation of hydrodynamic approximations  from kinetic equation and briefly outline  the way turbulence models appear from the NS equations.  In Section 3,  which is most important for the present development,  we introduce the non-equilibrium  kinetic equation and the zero-order pdf  for isotropic and homogeneous turbulence, both defined on an enlarged phase-space.   In  Section 4,  the finite energy flux is defined as a dynamic constraint and it is shown that the  introduced in Section 3 non-equilibrium pdf is indeed a solution to non-equilibrium  equation which conserves the total kinetic energy of a system.    Finally,  in  Section 5,   the coarse-grained kinetic equation for  hydrodynamic modes  is derived.  It is shown in  Sections 6 and 7,  that the CE expansion,   leads  to transport  equations  for  both  turbulent velocity field and kinetic energy.   In Section 8, choosing a proper relaxation time,   we show that our kinetic equation contains  the well known {\cal K}-{\cal E} , Reynolds stress and non-linear models.  In addition, this simple  equation contains turbulent models of an arbitrary non-linearity and complexity. Some new effects,  originating from   interaction of  spatial fluxes with the energy flux (cascade) are identified. 
Summary and conclusions are presented in Section 9.

\subsection{ Equilibrium.}  In equilibrium, with all temporal and spatial gradients equal to zero,   the left side of kinetic  equation (1.1),(1.2)  is equal to zero and the remaining equation ${\cal C}=0$ has a solution:

\begin{equation}
f^{eq}=constant\times e^{-\frac{E(\Gamma)}{\theta}}
\end{equation}
\noindent where the energy $E(\Gamma)$ is conserved  in the process of collision, i.e. $E'+E'_{1}=E+E_{1}$ and the temperature $\theta=const$. For a gas moving with the  constant velocity ${\bf V}$, the solution is:

\begin{equation}
f^{eq}=const \times e^{-\frac{E(\Gamma)-{\bf p\cdot V}}{\theta}}
\end{equation}
\noindent This solution is clear if we introduce internal energy $\epsilon_{in}$ and write

\begin{equation}
E(\Gamma)=\epsilon_{in}+\frac{mc^{2}}{2}
\end{equation}
\noindent  In  a frame of reference moving with velocity  ${\bf V}$, we have 

\begin{equation}
f^{eq}=constant\times e^{\frac{\mu-\epsilon_{in}}{\theta}}e^{-\frac{m({\bf c-V})^{2}}{2\theta}}
\end{equation}

\noindent In deriving transport approximation, one expands the probability density  powers the large-scale field (LL)
${\bf V}\approx {\bf V_{0}}+\frac{\partial {\bf V}}{\partial x_{i}}x_{i}$. The constant component $V_{0}=const$ can be removed by the Galileo transformation and only  derivatives of the large scale field ${\bf V}$ are relevant. 
 
\section{Hydrodynamic approximation}  
\noindent    
In this Section we, for the sake of clarity and continuity,  briefly describe the main steps leading from the Boltzmann equation to hydrodynamic approximations. 

\noindent Classic derivation of transport approximations starts with the  Boltzmann equation (1.1), (1.2) or (1.3) 
and equilibrium pdf 

\begin{equation}
f^{eq}=\frac{\rho}{(2\pi \theta)^{\frac{3}{2}}}\exp(-\frac{({\bf c-v}({\bf x},t))^{2}}{2\theta})
\end{equation}
 
\noindent  In this Section we denote  velocity of the frame of reference ${\bf v}({\bf x},t)$ which is a physical (not phase-space variable). In what follows we restore the notation  $({\bf V})$ used above.  This should not lead to any difficulty.
The equilibrium pdf is obtained from equation  ${\cal C}(f^{eq})=0$,  which is valid locally, so that the hydrodynamic field ${\bf v({\bf x},t})$ can be a  slow varying function of  time and space    coordinates.   In addition, the conservation laws and the symmetries of  collision integral give:

\begin{equation}
\int {\cal C}d{\bf  c}=\int {\cal C}(f){\bf c} d{\bf c}=0
\end{equation}

\noindent Introducing  the fluid density $\rho=\int f({\bf v}) d{\bf v}$ and integrating (1.1)-(1.2)  over ${\bf c}$  leads to the continuity equation:

\begin{equation}
\frac{\partial  \rho }{\partial t}+ \nabla \cdot \rho {\bf v}=0
\end{equation}

\noindent Multiplying (1.1)-(1.2) by ${\bf c}$ with subsequent integration over the phase space   yields :

\begin{equation}
\frac{\partial  \rho v_{j}}{\partial t}+ \partial_{i}  \rho v_{j}v_{i}+\partial_{i}  \sigma_{ij}=0
\end{equation}

\noindent where the stress tensor is :  $\sigma_{ij}=\overline{\rho(c_{i}-v_{i})(c_{j}-v_{j})}$. In thermodymanic equilibrium,   the stress-tensor is evaluated readily with the distribution function $f^{eq}$ from (2.1) giving $\sigma_{ij}|_{eq}=\rho\theta\delta_{ij}$,  so that  combining this  with the   ideal gas equation,  gives:
 
\begin{equation}
\frac{\partial  v_{j}}{\partial t}+ v_{i}\partial_{i} v_{j}-\frac{1}{\rho} \nabla p=0
\end{equation}

\noindent In what follows  $\overline{A}|_{eq}\equiv \int A f^{eq}d{\bf c}$. Defining $\theta=\frac{1}{d}\overline{(c_{i}-v_{i})^{2}}$, we obtain:

\begin{equation}
\frac{\partial \theta}{\partial t}+\nabla_{i}\rho v_{i}\theta+\frac{1}{d}\nabla_{i}\rho \overline{(c_{i}-v_{i})(c_{j}-v_{j})^{2}}
 +\frac{2}{d}\rho\overline{(c_{i}-v_{i})(c_{j}-v_{j})}S_{i,j}=0
\end{equation}

\noindent   
On the equilibrium pdf $f^{eq}$,  $\sigma^{eq}_{ij}=\theta\delta_{ij}$ and the third-order  in fluctuations contribution to  (2.6) is equal to zero. 

\subsection{Turbulence modeling.} The nature of the large-scale hydrodynamic field ${\bf v}$ in (2.4) has not been specified. The problem of turbulence modeling is formulated as follows:  let the velocity field ${\bf v}={\bf V}+{\bf v'}$
where $\overline{{\bf v'}}=0$ and $\overline{{\bf v}}={\bf V}$.  Our goal is to derive the equation for  the large-scale slow field ${\bf V}$. The above decomposition gives:

\begin{equation}
\frac{\partial {\bf V}}{\partial t}+{\bf V\cdot \nabla V}+\overline{{\bf v'\cdot \nabla v'}} +\frac{1}{\rho}\nabla\cdot  \sigma=0
\end{equation}
\noindent or 
 
\begin{equation}
\frac{\partial {\bf V}}{\partial t}+{\bf V\cdot \nabla V}+\frac{1}{\rho}\nabla\cdot ( \sigma+\sigma_{v})-\overline{{\bf v'\nabla  \cdot v'}}=0
\end{equation}

\noindent where 

\begin{equation}
\sigma_{ij,v}=\overline{(v-V)_{i}(v-V)_{j}}
\end{equation}
\noindent  
Now we need a theory or at least set of rules to express both stress $\sigma$ and  the Reynolds stress $\overline{{\bf v'\cdot \nabla v'}}$  in terms of the large-scale field ${\bf V}$ and its derivatives. This program has been systematically developed in Yakhot et al  (1992),  Smith et al  (1992).  The method consists of a few steps: 
1. In the first order,  the stress tensor $\sigma_{ij}\approx 2\nu_{0} S_{ij}$.  This leads to the Navier-Stokes equations (NS).
 2. ~
Add a statistically isotropic and homogeneous  stirring force  to  the  right side of the NS. 
This force generates an $O(1)$ energy flux ${\cal P}=\overline{{\bf f\cdot v'}}={\cal E}$ across scales.  In the case of a free turbulence with the total kinetic energy $K$, prepared as  an initial condition, the energy balance reads $\partial_{t}K=-{\cal E}$.  3.~ Use Wyld's (1961) expansion to eliminate small-scale velocity fluctuations.
It has been shown (Yakhot et al (1992)) that development of turbulent models  involves a double expansion in powers of   two dimensionless parameters: the Reynolds number $Re=v'_{rms}L/\nu$  and rate of strain $\eta_{ij}\equiv \eta=\tau S_{i,j}$ where $\tau_{T}$ is the relaxation time of turbulent fluctuation.  The expansion in Re leads to the fixed point   $Re_{*}\approx v'_{rms}L/\nu_{T}=const$, corresponding to the zero- spatial -gradient state of isotropic and homogeneous turbulence which serves a a ground state for the additional expansion in powers  $\eta$. The first expansion accounts for the non-zero energy flux of isotropic and homogeneous turbulence while the second~-~ gives rise to the not zero  mean velocity field and various fluxes.  In the first order of the $\eta$-expansion one derives a  variant of ${\cal K}-{\cal E}$ model, in the second -the Reynolds stress model and non-linear models etc.  It is important that   the second order $O(\nu_{T}^{2}/K {\bf S\cdot S})$ terms of the non-linear  turbulent models,  are well known from kinetic theory as next-order in the Chapman-Enskog expansion corrections to  the Navier-Stokes equations (Landau-Lifshitz (1981) ). This fact establishes similarities between Wyld's  expansion of the NS equations and the CE expansion of kinetic theory.
While, one can formally write each term of the $\eta$-expansion,  resummation of an infinite  series is totally out of reach.

\section { Formulation in terms of kinetic equation.}  In turbulent flows,   the velocity field is characterized by three correlation lengths (Landau and Lifshitz (1959), Frisch (1996)): 1. ~ microscopic  mean -free path $r\approx \lambda\approx c_{s}\tau$, where $c_{s}$ is the speed of sound and $\tau$ is the microscopic relaxation time;  2.~hydrodynamic inertial interval  $\eta_{K}\approx L Re^{-\frac{3}{4}}\leq r\leq L$, where  both  large-scale forcing  and viscous effects can be neglected and 3.~ nonuniversal,  geometry-dependent,   energy-production  range $r\geq L$. We denote three velocity fields corresponding to these intervals as 
   ${\bf c'\equiv c_{micro}= c-v; ~ v'=v-V; ~V}$, respectively.  (See Fig.1).  Locally,  at the scales $r\approx \lambda$,  the system is in thermodynamic equilibrium so that the ${\bf c_{micro}}$-field obeys Maxwell (Gibbs) statistics.

\begin{figure}\nonumber 
\centerline{\includegraphics[angle=0,scale=1, ,draft=false]{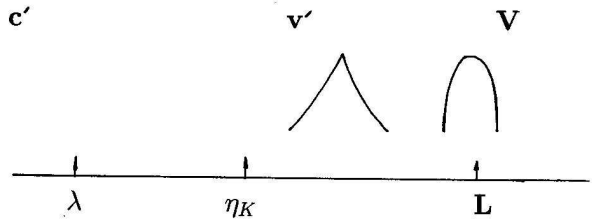}}
 \caption{ Scale   and velocity field intervals in turbulence:  1. ~Energy range  : $r >L$  ({\bf V} );  2.~Inertial range:  $\eta_{K}\leq r\leq  L$  ({\bf v'=v-V}). PDFs: $r\approx L$ (Gausss); $r\ll L$ (Multifractal);  $r\leq \lambda$ (Microscopic) ({\bf c'=c-v} ).}
\end{figure}

\noindent Hydrodynamic approximations  follow  elimination (small-scale averaging)  of  modes from the interval $r \leq \lambda\approx c_{s}\tau$,  where $\tau$ is the relaxation time characterizing dynamics of small perturbations from thermodynamic equilibrium.  After that, the turbulence models are derived by averaging the Navier-Stokes equations in the interval $\eta_{K}\leq r\leq L$.   We do not doubt the ability of the Navier-Stokes equations  fully describe turbulent flow of arbitrarily large Reynolds number.   In this case, since the number of excited modes is at least $O(Re^{9/4})$, the computational cost of  these direct simulations (DNS)  is very high.  Our concern here is with  derivation of the coarse-grained equations  (turbulence models) for partial representation of turbulence  in the interval $r\approx L$,  which is hard to obtain from hydrodynamics. In other words, 
{\it in this work we are interested in the mode-elimination from the interval $0 \leq r\leq L$ applied directly to the Boltzmann equation. If successful,  this way we will obtain the coarse-grained kinetic equation, containing, as a hydrodynamic approximation,  all possible turbulence models, which are very hard to derive from  the Navier-Stokes equation.}  
In particular, we are interested in equations of motion  for the coarse-grained or  filtered velocity field which includes  only a  small fraction of hydrodynamic modes corresponding to the top of  inertial range, i.e. $r\approx L$.  Whatever the filtering procedure,  the coarse-grained fields  are obtained from the full  fields  by averaging over  flow volumes of linear dimension $r\approx \Delta<L$, where $\Delta$ is the filtering scale or over the modes  ${\bf v(k)}$ with $k>1/\Delta$ in a Fourier space. Also, one can define a coarse - grained field as:

$${\bf v}({\bf x})\approx \frac{1}{{\cal R}({\bf x})}\int_{{\cal R}} {\bf v}({\bf x'})\ d{\bf x'}$$

\noindent where the integration is carried out over the volume ${\cal R}\approx \Delta^{3}$ centered about the point ${\bf x}$
\noindent or

$${\bf v}({\bf x})\approx \int  {\bf v}({\bf x'})F(|{\bf x-x'}|/\Delta)\ d{\bf x'}$$
\noindent where $F$ is a filter function. We see that the coarse-grained velocity field is a multipoint construction  and evaluation of high-order moments is not a simple task.  Moreover,    spatial velocity derivatives are defined on a cut-off as:  

$$\partial_{x} v=(v(x+\Delta_{x})-v(x))/\Delta_{x}$$ 

\noindent and,  if $\Delta\ll L$,   their moments,  called structure functions,  obey multifractal statistics. The  inertial range multifractality (anomalous scaling )  means  that the moments of different orders are independent of each other.  As a result,  the full statistical description of  velocity field involves a large, infinite in the limit $Re\rightarrow \infty$,  number of independent parameters.   This makes an accurate coarse-graining with the filter (mesh) size $\Delta<<L$ basically impossible. 

\noindent  The situation changes dramatically in the limit $\Delta\rightarrow L$.  It is a well-established, though not well-understood,  fact that at the scales $r=\Delta \approx L$,  both  the turbulent velocity fields  ${\bf v'=v-V}$ 
and $\delta {\bf v'}={\bf v'}({\bf x+\Delta})-{\bf v'}({\bf x})$  obey   close- to- gaussian statistics.  
Moreover,  it was demonstrated by Schumacher et al (2007) that  in the low-Reynolds number turbulence,  the velocity derivatives,  too, are  Gaussian random variables. If this is so,  the entire large-scale field  can  be accurately  described in terms of   the second-order moment only.   Our goal is to derive  a coarse - grained equation for  the  modes ${\bf v'}$ fluctuating  on the scales $r\approx  L$.  In the standard turbulence modeling,  this is  operationally achieved by introducing a large  ``effective viscosity''  $\nu_{T}\approx u_{rms}L$, where $2K=u^{2}_{rms}\approx \frac{3}{2}C_{K}L^{2}$,   into the Navier-Stokes equations. Due to a particular choice of viscosity $\nu_{T}\gg \nu_{0}$,  the small-scale modes ($r\ll L$) are {\it overdamped } and the resulting hydrodynamic equations  describe  large-scale properties ($r\approx L$) of the flow only.  The defined this way equation is called `` transport model''  as as opposed to the LES  models based on   an   inertial range  cut -off  $r=\Delta\ll L$.
Since the dominant  contribution to  turbulent kinetic energy is contained in  the large -scale fluctuations,  we write the zero-order nonequillibrium probability density  function in the laboratory reference frame:

\begin{equation}
f^{0}=\frac{\rho}{8\pi^{3}({\cal K}\theta)^{\frac{3}{2}}}\exp(-\frac{({\bf c-v})^{2}}{2\theta})\exp(-\frac{({\bf v-V})^{2}}{2{\cal K}})\equiv f^{0}({\bf v})Q^{0}({\bf c}|{\bf v})
\end{equation}

\noindent  
In this expression,  both ${\bf c}$ and ${\bf v}$- fields are phase-space variables independent of  time and space. Thus, we have temporarily  enlarged the phase space from six to nine dimensions. The flow described by the probability density function (3.1) can be perceived as an ensemble of laminar flows each in thermodynamic equilibrium with the {\it fluctuating equilibrium Maxwell-Gibbs probability density. } 
It is clear that: $\int f^{0}d{\bf v}d{\bf c}=\rho$ and  $\overline{\bf c}|_{0}\equiv \int c f^{0}d{\bf c}d{\bf v}={\bf V}$ and  $\overline{{\bf v}}|_{0}={\bf V}$.   
In the laminar limit  ${\cal K}\rightarrow 0$

$$f^{0}\propto \frac{1}{(2\pi\theta)^{\frac{3}{2}}}\exp(-\frac{({\bf c-v})^{2}}{2\theta})\delta({\bf v-V})$$

\noindent and integrating over ${\bf v}$, we recover the expression (2.1). The pdf (3.1),  resembling that of a mixture of gases with different temperatures $\theta$ and ${\cal K}$,  describes an essentially non-equilibrium state. {\it  It will be shown below that (3.1)  is a solution to  a ``nonequilibrium'' kinetic equation and this temperature difference is related to  the energy flux from hydrodynamic modes to the microscopic ones. }

\noindent From the above definitions:
\begin{eqnarray}
{\bf V}=\overline{\bf v}|_{0}=\int {\bf v}f^{0}({\bf c,v})d{\bf c}d{\bf v}\nonumber \\
\xi^{0}({\bf v})f^{0}({\bf v})=\int {\bf c}f^{0}({\bf c,v})d{\bf c}={\bf v}f^{0}({\bf v})\nonumber \\
{\cal K}=\frac{1}{d}\overline{({\bf v-V})^{2}}|_{0}=\frac{1}{d}\int ({\bf v-V)}^{2}f^{0}({\bf v, c})d{\bf c}d{\bf v}\nonumber \\
\end{eqnarray}

\noindent In what follows, we assume the probability density

\begin{equation}
f({\bf c, v, x},t)=\rho f(\frac{{\bf c-v}}{\sqrt{\theta}},\frac{{\bf v-V}}{\sqrt{{\cal K}}},t)
\end{equation}

\noindent so that  locally $\overline{({\bf c-v})^{2}}=d\theta$ and $\overline{({\bf V-v})^{2}}=d{\cal K}$ 

\noindent The pdf of hydrodynamic velocity field $f({\bf v})$ is obtained from (3.3) by integrating out the microscopic variable  ${\bf c}$, i.e.

\begin{equation}
f({\bf v})=\int f({\bf c,v,x},t)d{\bf c}
\end{equation}

\noindent  {\it Our goal is to derive a kinetic equation for  $f({\bf v})$  describing  turbulent fluctuations only. To achieve this goal, we first increase the original six-dimensional phase space of the previous section (${\bf c,x}$)  to the nine-dimensional one (${\bf c,v,x}$) and, eliminating thermodynamic fluctuations,  project the nine-dimensional problem to the six-dimensional one $({\bf v,x})$.  The advantage of this procedure from the turbulence modeling viewpoint is that while  original problem involved $N_{\lambda}\approx \Omega/\lambda^{3}$ number of particles, the coarse-grained one ~-~ only $N_{L}\approx \Omega/\L^{3}$,  where the integral scale $L\gg\lambda $ is the large scale correlation length of turbulence. }

\noindent Since the fast ${\bf c_{micro}= c-v}$  -field varies on the time scale $\tau\approx \lambda/c_{s}$, where $c_{s}$ stands for the speed of sound,   $\overline{{\bf (c-v)\cdot v}}=0$, we can consider these components statistically ortogonal and introduce the continuity equation in the enlarged  9d-phase- space.  In this case the Botzmann equation (Landau/Lifshitz (1981)):

\begin{equation}
\frac{\partial f}{\partial t}+{\bf c}\cdot \nabla f= I(f({\bf c,v}))
\end{equation}

\noindent where $I(f)$ is not yet specified `non-equilibrium collision integral'  with $f^{0}({\bf v,c,x},t)$ obtained from equation:

\begin{equation}
\frac{\partial f^{0}}{\partial t}= I(f^{0}({\bf c,v}))\neq 0
\end{equation}

\noindent  with $V=const$ ($S_{ij}=0)$.  {\it Thus, in the theory developed below, we abandon the equilibrium  pdf   (2.1)  for the the zero-spatial-gradient  state of a gas   in favor of the nonequilibrium expression (3.1). } The detailed discussion of this point will be given below.  Here we just mention that  due to the non-zero energy flux across the scales,  the kinetic energy of the zero-spatial-gradient  ($\eta=0$) turbulent flow decays leading to increase of the gas temperature $\theta$.  
 
\section{Finite energy flux as a dynamic constraint.}  

\noindent    If a turbulent flow is supported by  a  large-scale energy input per second ${\cal P}$,  by nonlinearity, this energy is transfered  to the smallest scales where it is dissipated. 
 It is usually assumed that the nature of the dissipation mechanism is unimportant.   On  the inertial- range scales, where both energy pumping and  dissipation are negligibly small, the dynamics are characterized by the mean  energy flux 
$J={\cal P}={\cal E}$,  where ${\cal E}$  is the mean dissipation rate.  In case of isotropic and homogeneous turbulence,  prepared at initial instant of time with kinetic energy $K(0)$,   the energy  $K(t)$ decays to zero. Since in this case,  all spatial gradients are equal to zero,   no {\it mean spatial fluxes } are involved in the relaxation process. 
{\it Unlike equilibrium gas,   this strongly non-equilibrium system is characterized by a non-zero constant energy flux {\it across the scales }}.

\noindent  It has been shown in a series of extraordinary works by Zakharov et al (1975), (1992), (1984)   that, in case of turbulence,  the Euler equations,  with velocity field written  in terms of Clebsch variables, can be represented as a kinetic equation for  waves (particles) with the collision  integral $I_{C}$ describing the non-linear  interactions. 
It has been shown that,  in addition to equilibrium solutions,  the equation  $I_{C}(n({\bf k}))=0$ for the occupation numbers $n(k)$,  playing the role of the pdf in the $k$-space,    has  solutions corresponding to the non-zero constant fluxes in the wave-number space.  The direction of these fluxes is exclusively determined by physics of the problem and peculiarities of  non-linear interactions.   For example,  the mean field approximation developed in Yakhot (1992),   led to  the Kolomogorov energy spectrum as  one of  Zakharov's constant-flux solutions.  Thus, unlike thermodynamic equilibrium,  the kinetic equation for  a zero -mean-gradient turbulent flow  must satisfy the  flux (energy balance) constraints:
 
 \begin{equation}
 {\cal P}={\cal E}
 \end{equation}
\noindent valid in the statistically steady state or 

\begin{equation}
\frac{\partial {\cal K}}{\partial t}=-\frac{2}{d}{\cal E}
\end{equation}
\ - \ in case of decaying turbulence.

\noindent 

\noindent The relations (4.2) and (4.3),  reflecting the energy balance in a flow,   are independent upon  the nature of the dissipation mechanism.   
\noindent  We split collision integral $I(f)$ into two components:  $I_{hi}(f)$  responsible for the non-zero energy flux of isotropic and homogeneous turbulence and $I^{1}(f)$~-~ for the spatial fluxes of a general turbulent flow.  Since, as was pointed out above,  the mean energy flux is a large-scale property, no differential operator can enter the expression.  Thus we have: $I(f)=I_{hi}+I^{1}$ and 

\begin{equation}
I_{hi}=-\frac{f}{\tau_{hi}}=-[\frac{({\bf v-V})^{2}-d{\cal K}}{d{\cal K}}-\frac{{\cal K}}{\theta} \frac{({\bf v-c})^{2}-d\theta}{d\theta}]\frac{{\cal E}}{{\cal K}}f({\bf c,v})
\end{equation}

\subsection{ Isotropic and homogeneous turbulence.}  

\noindent Now we will show that the zero-order pdf (3.1) is a solution to kinetic equation describing isotropic and homogeneous turbulence. Since in this case, all spatial derivatives are equal to zero,  the kinetic equation  is:

\begin{equation}
\frac{\partial f^{0}}{\partial t}=I_{hi}
\end{equation}

\noindent and $\overline{{\bf c}}=\overline{{\bf v}}=0$, $\overline{c^{2}}=d\theta$ and $\overline{v^{2}}=d{\cal K}$. Substituting (3.1) into (3.5)-(3.6) gives:

\begin{equation}
\frac{1}{\theta}(\frac{({\bf v-c})^{2}}{\theta}-d)(\frac{\partial \theta}{\partial t}-\frac{2}{d}{\cal E})=\frac{1}{{\cal K}}(\frac{v^{2}}{{\cal K}}-d)(\frac{\partial {\cal K}}{\partial t}+\frac{2}{d}{\cal E})
\end{equation}

\noindent This equation, valid for  all values of independent parameters $c$ and $v$, can be correct only if:

\begin{equation}
\frac{\partial K}{\partial t}=-{\cal E}
\end{equation}

\begin{equation}
\frac{d}{2}\frac{\partial \theta}{\partial t}={\cal E}
\end{equation}

\noindent  {\it Thus,  we conclude that the zero-order pdf (3.1) is indeed a solution to the kinetic equation (3.5), (3.6) for isotropic and homogeneous turbulence, subject to the energy constraints (4.1),(4.2).
One can also see that  according to (4.6),  (4.7): $\frac{d}{2}({\cal K}+\theta)\equiv K+T=const$, which reflects conservation of the total kinetic  energy in decaying isotropic and homogeneous  turbulence.}

\section{ Coarse-graining.}
 Introducing conditional probability density $Q({\bf c}|{\bf v})$ we can write: 

\begin{equation}
f({\bf v},{\bf c},{\bf x},t)=f({\bf v},{\bf x},t)Q({\bf c}|{\bf v})
\end{equation}

\noindent so that 

\begin{equation}
f({\bf v},{\bf x},t)=\int f({\bf v},{\bf x},t)Q({\bf c}|{\bf v})d{\bf c}
\end{equation}
\noindent 

\begin{equation}
\int c_{\alpha}Q({\bf c}|{\bf v})d{\bf c}=\xi_{\alpha}({\bf v})+v_{\alpha}
\end{equation}

\noindent  and 
\begin{equation}
\int d{\bf v} \xi_{\alpha}({\bf v}) f({\bf v})=V_{\alpha}({\bf x})
\end{equation}

\noindent  Below we develop an approach  resembling the one proposed for the probability densities of the passive scalar in a random velocity field in Sinai, Yakhot (1989).   Integrating the nine-dimensional Boltzmann  equation (3.5), (4.3) over the relatively fast- field ${\bf c}$ gives:

\begin{equation}
\frac{\partial f({\bf v})}{\partial t}+  \nabla \cdot {\bf v}f({\bf v})+\nabla\cdot \xi({\bf v})f({\bf v})= {\cal I}(f)+i^{1}
\end{equation}

\noindent where

\begin{equation}
{\cal I}=-\frac{({\bf v-V})^{2}-d{\cal K}}{d{\cal K}} \frac{{\cal E}}{{\cal K}}f({\bf v})
\end{equation}
\noindent and the  collision integral 

$$
i^{1}=-\frac{f-f^{0}({\bf v})}{\tau_{T}}
$$

\noindent is formally written with the yet unspecified relaxation time $\tau_{T}$. The coarse -grained zero-order pdf is:

\begin{equation}
f^{0}({\bf v})=\frac{\rho}{(2\pi {\cal K})^{\frac{d}{2}}}e^{-\frac{({\bf v-V})^{2}}{{2{\cal K}}}}
\end{equation}

\subsection{Model for conditional expectation value.}  The expression for conditional expectation value $\xi_{i}({\bf v})$ is a subject to four dynamic constraints.  

\noindent 1.~ It must be invariant under  Galileo transformation and not violate the Galileo invariance of equation (5.5).  This is the reason for the $v_{\alpha}$ term in the right side of (5.3).

\noindent  2.~ Integrating (5.5) over ${\bf v}$ must lead to the continuity equation. This gives:

\begin{equation}
\nabla\cdot \int \xi({\bf v})f({\bf v}) d{\bf v}=0
\end{equation}
\noindent 
\noindent 3.~ Multiplying (5.5) by ${\bf v}$ and comparing the result with (2.6) yields: 

\begin{equation}
\nabla_{i} \int  v_{j}\xi_{i}({\bf v})f^{0}({\bf v}) d{\bf v}=\nabla_{i} \overline{\rho (c_{i}-v_{i})(c_{j}-v_{j})}|_{0}=\nabla_{j}\rho\theta
\end{equation}

\noindent  4.~ In the zeroth order of expansion in powers of dimensionless spatial gradients (Chapman-Enskog expansion), the equation (5.5) must generate  (2.5)  with the pressure term containing  information  about the integrated out microscopic variables.  In a simplest case of ideal gas, $p'=\rho\theta$.

\noindent A relation satisfying the  above constraints is:

\begin{equation}
\xi({\bf v})=\frac{({\bf v-V})}{{\cal K}}\cdot \nabla p'(\theta)
\end{equation}

\noindent where

$$\delta_{ij}p'(\theta)/\rho=\delta_{ij}\theta=\frac{\theta}{{\cal K}}\int (v-V)_{i}(v-V)_{j}f^{0}({\bf v})d{\bf v}$$

\noindent  Thus, the  course-grained kinetic  equation, correctly describing  isotropic and homogeneous turbulence (zero spatial gradients) and satisfying the dynamic constraints presented above, is :

\begin{equation}
\frac{\partial f({\bf v})}{\partial t}+  \nabla \cdot {\bf v}f({\bf v})+
\frac{{\bf v-V}}{{\cal K}}\cdot (\nabla p')f({\bf v})+\frac{({\bf v-V})^{2}-d{\cal K}}{{\cal K}}\frac{{\cal E}}{d{\cal K}}f({\bf v})=i^{1}
\end{equation}

\noindent The equation (5.11) is the main result of this paper.

\subsection{Hydrodynamic approximation .}

\noindent Integrating  (5.11) over ${\bf v}$ gives continuity equation (2.3).  In addition, multiplying it by ${\bf v}$  and integrating leads to:

\begin{equation}
\frac{\partial \rho V_{j}}{\partial t}+ \nabla_{i}[ \rho ( V_{i}V_{j} +\overline{(v_{i}-V_{i})(v_{j}-V_{j})})]+\frac{\overline{(v_{i}-V_{i})(v_{j}-V_{j})}}{{\cal K}}\frac{\partial p'}{\partial x_{i}}
+\frac{ \overline{\rho({\bf v-V})({\bf v-V})^{2}}}{{\cal K}}\frac{{\cal E}}{d{\cal K}}=0
\end{equation}
\noindent or 

\begin{equation}
\frac{\partial {\bf V}}{\partial t}+{\bf V\cdot \nabla V}+\frac{1}{\rho}\nabla_{i}[\overline{\rho (v_{i}-V_{i})(v_{j}-V_{j})}]+\frac{\overline{(v_{i}-V_{i})(v_{j}-V_{j})}}{\rho{\cal K}}\frac{\partial p'}{\partial x_{i}}+
\frac{\overline{\rho({\bf v-V})({\bf v-V})^{2}}}{\rho {\cal K}}\frac{{\cal E}}{d{\cal K}}=0
\end{equation}
\noindent  
\noindent In the zeroth order, averaged with  with the pdf (5.7), the equation (5.14) gives  the Euler equation:

\begin{equation}
\frac{\partial {\bf V}}{\partial t}+{\bf V\cdot\nabla V}=-\frac{1}{\rho}\nabla p
\end{equation}

\noindent   where $p=\rho(\theta+{\cal K})$.   It follows from (5.7) and (5.11) that in the case of strongly nonequilibrium flow as an unperturbed state, the equation of state $p'(\theta,\rho)$  in (5.11) is decoupled from the ``temperature'' ${\cal K}$ entering the problem as a parameter in the zero -order pdf (5.7).  Indeed, the  coarse grained equation (5.11) gives for $\frac{1}{d}\overline{{\bf ( v-V)^{2}}} ={\cal K}$ :

\begin{eqnarray}
\frac{\partial \rho{\cal K}}{\partial t}+\nabla_{i}(\rho V_{i}{\cal K})+ \frac{2}{d}\rho \overline{(v_{i}-V_{i})(v_{j}-V_{j})}S_{ij}+\frac{1}{d}\nabla_{i}\overline{\rho (v_{i}-V_{i})(v_{j}-V_{j})^{2}}+\frac{2}{d}\rho{\cal E} \nonumber \\
+\frac{1}{d}\frac{\overline{{\bf (v-V)(v-V)^{2}}}}{{\cal K}}\cdot \nabla p'
=0
\end{eqnarray}

\noindent  In the zeroth order ($\eta=0$),  all gradients equal to zero, the equation (5.15) reads:

\begin{equation}
\frac{\partial {\cal K}}{\partial t}=-\frac{2}{d}{\cal E}
\end{equation}
\noindent 

\noindent We also  define a vector 

$$\Sigma_{i}=\overline{(v_{i}-V_{i})({\bf v-V})^{2}}$$

\section{ Expansion.}

To find the solution to the kinetic equation (5.11) we following Chen et al (2004) write:

\begin{equation}
f({\bf v})=f^{0}+\epsilon f^{(1)}+\epsilon^{2}f^{(2)}+\cdot\cdot\cdot
\end{equation}

\noindent and 

\begin{equation}
\frac{\partial }{\partial t}=\epsilon \frac{\partial }{\partial t_{0}}+\epsilon^{2}\frac{\partial}{\partial t_{1}}\epsilon^{3}\frac{\partial}{\partial t_{3}}+\cdot\cdot \cdot
\end{equation}

\noindent and $\nabla=\epsilon \nabla_{1}$. The mean of any flow property is then 

\begin{equation}
\overline{A}=A^{(0)}+A^{(1)}+A^{(2)}+\cdot\cdot
\end{equation}

\noindent  where 

\begin{equation}
A^{(n)}=\int A({\bf v})f^{(n)} d{\bf v}
\end{equation}
\noindent 
The zero-order pdf $f^{(0)}$ is given by expression (5.7).
The first and second -order corrections  are  calculated readily with 
$i^{1}=-\frac{f-f^{0}}{\tau_{T}} $.
\noindent  The result is:

\begin{equation}
B[f^{(0)}({\bf v})]\equiv [\frac{\partial }{\partial t}+{\bf v}\cdot \nabla +  \frac{{\bf v-V}}{{\cal K}}\cdot(\nabla p')+\frac{({\bf v-V})^{2}-d{\cal K}}{{\cal K}}\frac{{\cal E}}{d{\cal K}}]f^{(0)}=-\frac{f^{(1)}}{\tau_{T}}
\end{equation}
\noindent and

\begin{eqnarray}
-B(\tau_{T}B f^{(0)}({\bf v}))+\frac{\partial f^{(0)}}{\partial t_{1}}  \equiv 
 [\frac{\partial }{\partial t_{0}}+  \nabla \cdot {\bf v}+ \frac{{\bf v-V}}{{\cal K}}\cdot (\nabla p')+\frac{({\bf v-V})^{2}-d{\cal K}}{{\cal K}}\frac{{\cal E}}{d{\cal K}}]f^{(1)}-\frac{\partial f^{(0)}}{\partial t_{1}}=-\frac{f^{(2)}}{\tau_{T}}
\end{eqnarray}

 \noindent This gives:

\begin{eqnarray}
-\frac{f^{(1)}}{\tau_{T}}=\frac{f^{(0)}}{2{\cal K}}[2(v-V)_{\alpha}(v-V)_{\beta}(S_{\alpha\beta}-\frac{\delta_{\alpha,\beta}}{d}S_{pp})+(\frac{({\bf v-V})^{2}}{{\cal K}}-d)(v-V)_{\alpha}\frac{\partial {\cal K}}{\partial x_{\alpha}}) 
\nonumber \\ -(\frac{({\bf v-V})^{2}}{{\cal K}}-d)\frac{{\cal K}}{d}\nabla\cdot  {\bf V}
-2(v-V)_{\alpha}\frac{\partial {\cal K}}{\partial x_{\alpha}}]+\frac{f^{0}}{\rho}({\bf (v-V)\cdot \nabla \rho}-\rho {\bf \nabla \cdot V})~~~~~
\end{eqnarray}

 \noindent and 
  
\begin{eqnarray}
-\frac{f^{(1)}}{\tau_{T}}=\frac{f^{(0)}}{2{\cal K}}[2(v-V)_{\alpha}(v-V)_{\beta}(S_{\alpha\beta}-\frac{\delta_{\alpha,\beta}}{d}S_{pp})+(\frac{({\bf v-V})^{2}}{{\cal K}}-d-2)(v-V)_{\alpha}\frac{\partial {\cal K}}{\partial x_{\alpha}}) 
\nonumber \\ 
 -2(v-V)_{\alpha}\frac {\partial \theta}{\partial x_{\alpha}} -(\frac{({\bf v-V})^{2}}{{\cal K}}-d)\frac{{\cal K}}{d}\nabla\cdot  {\bf V}]+
\frac{f^{0}}{\rho}({\bf (v-V)\cdot \nabla \rho}-\rho {\bf \nabla \cdot V})~~~~~
\end{eqnarray}

\subsection{First order. Momentum equation.}

\noindent Using the pdf  given by (6.8),  we can write:

\begin{equation}
\sigma_{ij}=\sigma_{ij}^{(0)}+\sigma_{ij}^{(1)}=\rho ({\cal K}+p'(\theta))\delta_{ij}-2\rho\tau_{T}{\cal K}(S_{ij}-\frac{\delta_{ij}}{d}S_{pp})-\tau_{T}\rho{\cal K}\nabla\cdot V\delta_{ij}
\end{equation}

\noindent  and the normal stress (pressure)   $p={\cal K}+p'$. Setting $\rho\theta=const$ gives;

\begin{equation}
\Sigma=\Sigma^{0}+\Sigma^{1}=-(d+2)\frac{\tau_{T}{\cal E}}{\rho{d{\cal K}}}\nabla(\rho{\cal K}) 
\end{equation}

\noindent  and:

\begin{eqnarray}
\frac{\partial V_{j}}{\partial t}+V_{i} \nabla_{i} V_{j} +\frac{1}{\rho}\nabla_{j} p -(d+2)\frac{\tau_{T}{\cal E}}{\rho d{\cal K}}\nabla_{j}\rho{\cal K}
=
\nabla_{j}{\cal K}\tau_{T}(2 S_{ij}+\nabla\cdot {\bf V})
\end{eqnarray}

\noindent or introducing turbulence kinetic energy $K=\frac{d}{2}{\cal K}$:

\begin{eqnarray}
\frac{\partial V_{j}}{\partial t}+V_{i} \nabla_{i} V_{j} +\frac{1}{\rho}\nabla_{j} p -(d+2)\frac{\tau_{T}{\cal E}}{\rho K}\nabla_{j}\rho K=
\frac{2}{d}\nabla_{j} K\tau_{T}(2 S_{ij}-\delta_{ij}\nabla\cdot {\bf V})
\end{eqnarray}

\noindent where $p=\rho(\theta+{\cal K})$. 

\noindent Two remarks are in order. Firsrt, an additional  dimensionless expansion parameter,   $R_{*}=\tau_{T}{\cal E}/{\cal K}$, 
 which is not present in the ordinary Chapman-Enskog expansion,  appears in the present nonequilibrium theory.  In strong turbulence,  where  molecular relaxation can be neglected,  the characteristic time is $\tau_{T}\approx {\cal K}/{\cal E}$ and  $R_{*}=O(1)=const$, is nothing but the renormalized Reynolds number $R_{*}=v'_{rms}L/\nu_{T}=O(1)$,  corresponding to the fixed point of isotropic and homogeneous turbulence.  Second, the derived pressure  $p=\rho({\cal K}+\theta)$ leads to renormalization of the sound speed  by the turbulent velocity fluctuations,  derived in Staroselsky et al (1990) and predicted earlier by Candrasekhar (1995) in the context of gravitational collapse of interstellar gas.

\subsection{First order. Energy equation.}  

\noindent Combining the above results we,  writing for the sake of simplicity of notation $S_{ij}-\frac{\delta_{ij}}{d}S_{pp}\equiv S_{ij}$,  have:

\begin{equation}
\frac{\partial {\cal K}}{\partial t}+{\bf V\cdot \nabla}{\cal K}=\frac{4}{d}\tau_{T}{\cal K}S_{ij}^{2}-\frac{2}{d}{\cal E}+\Sigma^{(1)}\cdot \nabla p'+ \frac{(d+2)}{d}\nabla\cdot {\cal K}\tau_{T}\nabla {\cal K}+\frac{2{\cal K}}{d}\nabla\cdot V
\end{equation}

 \noindent  In what follows the last term in (6.15), responsible for the so called dilatation effects in compressible flows, will be neglected. Multiplying this equation by $d/2$  we summarize the above results: {\it In the first order of the Chapman -Enskog expansion,
 the coarse grained kinetic equation (5.11) gives}  :

\begin{eqnarray}
\frac{\partial V_{j}}{\partial t}+V_{i} \nabla_{i} V_{j} +\frac{1}{\rho}\nabla_{j} p -(d+2)\frac{\tau_{T}{\cal E}}{\rho K}\nabla_{j}\rho K
=\nonumber \\
\frac{2}{d}\nabla_{j} K\tau_{T}(2 S_{ij}+\delta_{ij}\nabla\cdot {\bf V})
\end{eqnarray}

 \noindent and

\begin{equation}
\frac{\partial K}{\partial t}+{\bf V\cdot }\nabla K=\frac{4}{d}\tau_{T}KS_{ij}^{2}-{\cal E}+\frac{d}{2}\Sigma^{(1)}\cdot \nabla p'+ \frac{2(d+2)}{d^{2}}\nabla\cdot K \tau_{T}\nabla K
\end{equation}

\section{Second Order.}

\noindent  
According to (6.6)

\begin{equation}
f^{(2)}=\tau_{T}B\tau_{T}Bf^{(0}-\tau_{T}\frac{\partial f^{(0})}{\partial t_{1}}
\end{equation}

\noindent where the operator $B$ is defined in (6.5).  The symbol $\partial_{t_{1}}$ stands for the time-derivative in  first-order contributions to the hydrodynamic equations. For example:

\begin{equation}
\frac{\partial \rho}{\partial t_{1}}=0
\end{equation}

\begin{eqnarray}
\frac{\partial V_{j}}{\partial t_{1}}= (d+2)\frac{\tau_{T}{\cal E}}{\rho d{\cal K}}\nabla_{j}\rho{\cal K}+
\nabla_{j}{\cal K}\tau_{T}(2 S_{ij}+\delta_{ij}\nabla\cdot {\bf V})
\end{eqnarray}

\noindent and

\begin{equation}
\frac{\partial {\cal K}}{\partial t_{1}}=\frac{4}{d}\tau_{T}{\cal K} S_{ij}^{2}+ \frac{(d+2)}{d}\nabla\cdot \tau_{T}{\cal K}\nabla  {\cal K}-\frac{(d+2)\tau_{T}{\cal E}}{\rho d {\cal K}}(\nabla p')\cdot (\nabla  {\cal K})
\end{equation}

\noindent  It is clear that $\frac{\partial}{\partial t_{1}}$ is a time derivative of all $O(\tau_{T})$ contributions to expansion and, according to (7.1),  $f^{(2)}=O(\tau_{T}^{2})$.  Thus,

\begin{eqnarray}
 \frac{\partial f^{(0)}}{\partial t_{1}}=\frac{f^{(0)}}{2{\cal K}}[(\frac{({\bf v-V})^{2}}{{\cal K}}-d) (\frac{4}{d}\tau_{T}{\cal K} S_{ij}^{2}+ \frac{(d+2)}{d}\nabla\cdot \tau_{T}{\cal K}\nabla  {\cal K} -\frac{(d+2)\tau_{T}{\cal E}}{\rho d {\cal K}}(\nabla p')\cdot (\nabla  {\cal K}
)
 -\nonumber \\  \frac{2(v_{i}-V_{i})}{\cal K}( 2(d+2)\frac{\tau_{T}{\cal E}}{\rho d{\cal K}}\nabla_{i}\rho{\cal K}+
\nabla_{j}{\cal K}\tau_{T}(2 S_{ij}+\nabla_{j}\cdot {\bf V}) )]~~~~~~~~
\end{eqnarray}

\noindent Now, assuming $\rho=const$, $\theta=const$ and $\nabla\cdot {\bf V}=0$,  we rewrite (6.8) as

$$f^{(1)}=-\frac{\tau_{T}f^{(0)}}{2{\cal K}}G$$

\noindent where 

$$     G=[2(v-V)_{\alpha}(v-V)_{\beta}(S_{\alpha\beta}-\frac{\delta_{\alpha,\beta}}{d}S_{pp})+(\frac{({\bf v-V})^{2}}{{\cal K}}-d-2)(v-V)_{\alpha}\frac{\partial {\cal K}}{\partial x_{\alpha}} ]       $$

\noindent and 

$$f^{(2)}=\tau_{T}B\tau_{T}Bf^{(0)}- \tau_{T}\frac{\partial f^{(0)}}{\partial t_{1}}=\frac{f^{(0)}}{4{\cal K}^{2}}\tau_{T}^{2}G^{2}+\tau_{T}f^{(0)}(\frac{\partial}{\partial t_{0}}+{\bf v\cdot\nabla})(\frac{\tau_{T}G}{2{\cal K}})-
 \frac{\partial f^{(0)}}{\partial t_{1}}$$

\noindent Thus:


\begin{eqnarray}
Bf^{(1)}=-B\tau_{T}Bf^{(0)}=\hspace{3in}\nonumber \\
-\frac{f^{(0)}}{4{\cal K}^{2}}\{\tau_{T}[2(v-V)_{\alpha}(v-V)_{\beta}S_{\alpha\beta}+(\frac{({\bf v-V})^{2}}{{\cal K}}-d-2)(v-V)_{\alpha}\frac{\partial {\cal K}}{\partial x_{\alpha}}]^{2} +\hspace{1in}
\nonumber \\ 
 2{\cal K}\tau_{T}[-2[(v-V)_{\alpha}(v-V)_{\beta})-
  (v-V)_{\beta}\nabla_{\alpha}p-
   (v-V)_{\alpha}\nabla_{\beta} p]S_{\alpha,\beta})+
 \frac{1}{\tau_{T}}(v-V)_{\alpha}(v-V)_{\beta}({\cal D}\tau_{T}S_{\alpha\beta}  )  ]\}\nonumber \\
 -f^{(0)}\frac{\tau_{T}G}{2{\cal K}^{2}} ({\bf v-V)\cdot \nabla}{\cal K}\hspace{3in}
\end{eqnarray}

\noindent where $({\cal D}A)\equiv (\frac{\partial }{\partial t}+{\bf v}\cdot \nabla )A$.  
 Collecting all contributions up to $O((\nabla {\cal K})^{2})$ simple calculation  with   $\xi_{i}\equiv (v-V)_{i}$ gives:
 
 \begin{eqnarray}
 \frac{f^{(2)}}{\tau_{T}}=\frac{\tau_{T}f^{0}}{{\cal K}^{2}}
 \{
 \xi_{\alpha}\xi_{\beta}\xi_{\gamma}\xi_{\delta}S_{\alpha,\beta}S_{\gamma,\delta}+{\cal K}[-((\xi_{\beta}(\xi_{p}\nabla_{p} V_{\alpha} +\nabla_{\alpha} p)+\xi_{\alpha}(\xi_{p}\nabla_{p}V_{\beta} +\nabla_{\beta} p))S_{\alpha,\beta} + \nonumber \\
(S_{\alpha,\beta}\nabla_{\gamma}{\cal K})\xi_{\alpha}\xi_{\beta}\xi_{\gamma}(\frac{\xi^{2}}{{\cal K}}-d-1)+\frac{1}{\tau_{T}} \xi_{\alpha}\xi_{\beta}({\cal D}\tau_{T}S_{\alpha,\beta})\}
  -\frac{\partial f^{(0)}}{\partial t_{1} }+O((\nabla {\cal K})^{2})\hspace{1in}
\end{eqnarray}
 
\noindent so that  for $i\neq j$:
 
 \begin{equation}
 \sigma_{ij}^{(2)}=\frac{\tau_{T}^{2}{\cal K}^{2}}{{\cal K}}\{8S_{ip}S_{pj}-2\frac{\partial V_{\alpha}}{\partial x_{j}}S_{\alpha,i}-2\frac{\partial V_{\alpha}}{\partial x_{i}}S_{\alpha,j}+
\frac{2}{\tau}({\cal D} \tau_{T}S_{ij} )\} +O(\nabla {\cal K})
 \end{equation}
 
\noindent and finally:
 
 \begin{equation}
\sigma^{(2)}= \frac{\tau_{T}^{2}{\cal K}^{2}}{{\cal K}}[2\frac{\partial V_{i}}{\partial x_{\alpha}}\frac{\partial V_{j}}{\partial x_{\alpha}}+\frac{\partial V_{j}}{\partial x_{\alpha}}\frac{\partial V_{\alpha}}{\partial x_{i}}
+\frac{\partial V_{i}}{\partial x_{\alpha}}\frac{\partial V_{\alpha}}{\partial x_{j}}+\frac{2}{\tau_{T}}({\cal D}\tau_{T}S_{ij})] +O(\nabla {\cal K})
\end{equation}
 
\noindent The remaining $O(\nabla {\cal K})$ contributions can be easily calculated. For example  :

\begin{equation}
\Sigma_{i}^{(2)}=\overline{\xi_{i}\xi_{\alpha}\xi_{\beta}\xi_{\gamma}\xi^{2}
(\frac{\xi^{2}}{{\cal K}}-d-1)}|_{0} \frac{\tau_{T}^{2}{\cal E}}{d{\cal K}^{4}}S_{\alpha,\beta}\nabla_{\gamma}{\cal K}
=\frac{2(d+3)(d+5)}{d}\frac{\tau_{T}^{2}{\cal E}}{{\cal K}}S_{i,j}\nabla_{j}{\cal K}
\end{equation}
 
\noindent We see that  

\begin{equation}
\frac{\Sigma_{i}^{(2)}}{\Sigma_{i}^{(1)}}\approx \frac{2(d+3)(d+5)}{d(d+2)}\tau_{T}|S_{ij}|
\end{equation}

\noindent The usually neglected $O(\nabla K)$ contributions to the turbulent models, may not be too small in the high-gradient regions of the flow.

\section{Turbulence Models Generator.} 

\noindent Introducing effective viscosity 

$$\nu({\cal K}, {\cal E}, |S_{ij}|, x,t)=\nu_{0}+\nu_{T}$$

\noindent with 

\begin{equation}
\nu_{T}=\frac{2}{d}\tau_{T}\frac{d}{2}{\cal K}\equiv \frac{2}{d}\tau_{T}K
\end{equation}

\noindent gives:

\begin{equation}
\sigma_{ij}=({\cal K}+p'(\theta))\delta_{ij}-2\nu S_{ij}+2\nu( {\cal D} \tau_{T}S_{ij})+\frac{d}{2}\frac{\nu^{2}}{K}[2\frac{\partial V_{i}}{\partial x_{\alpha}}\frac{\partial V_{j}}{\partial x_{\alpha}}+\frac{\partial V_{j}}{\partial x_{\alpha}}\frac{\partial V_{\alpha}}{\partial x_{i}}
+\frac{\partial V_{i}}{\partial x_{\alpha}}\frac{\partial V_{\alpha}}{\partial x_{j}}] +O(\nabla {\cal K})
\end{equation}

\noindent 
 We can summarize the results of this section:   in the second order of expansion if $\rho\theta=const$ and ${\bf \nabla\cdot V}=0$,  the turbulence model reads:

\begin{eqnarray}
\frac{\partial V_{j}}{\partial t}+V_{i} \nabla_{i} V_{j} +\frac{1}{\rho}\nabla_{j} p -(d+2)\frac{\tau_{T}{\cal E}}{\rho K}\nabla_{j}\rho K
=\nonumber \\
2\nabla_{i}\nu S_{ij}  +\nabla_{i}\sigma^{(2)}_{ij}-\frac{2(d+3)(d+5)}{d}\frac{\tau_{T}^{2}{\cal E}}{{\cal K}}S_{i,j}\nabla_{i}{\cal K}
\end{eqnarray}

 \noindent and

\begin{equation}
\frac{\partial K}{\partial t}+{\bf V\cdot }\nabla K=\frac{4}{d}\tau_{T}KS_{ij}^{2}-{\cal E}+\frac{d}{2}\Sigma^{(1)}\cdot \nabla p'+ \frac{2(d+2)}{d^{2}}\nabla\cdot K \tau_{T}\nabla K-\nabla_{i}\frac{2(d+3)(d+5)}{d^{2}}\frac{\tau_{T}^{2}{\cal E}}{K}S_{i,j}\nabla_{j}K
\end{equation}
 
 \noindent In the high Reynolds number regions of the flow, $R_{*}=\tau_{T}{\cal E}/K=const$ and

\begin{equation}
\frac{\partial K}{\partial t}+{\bf V\cdot }\nabla K=\frac{4}{d}\tau_{T}KS_{ij}^{2}-{\cal E}+\frac{d}{2}\Sigma^{(1)}\cdot \nabla p'+ \frac{2(d+2)}{d^{2}}\nabla\cdot K \tau_{T}\nabla K-\frac{2(d+3)(d+5)}{d^{2}}R_{*}S_{i,j}\nabla_{i}(\tau_{T}\nabla_{j}K)
\end{equation}

\noindent The expressions (8.4) and (8.3) have a few features not present in  familiar turbulence models.   First, consider a simplest possible example of the fully developed  statistically steady flow in an infinite channel. In this case,  $\sigma^{(2)}=0$, $\partial_{t} u=\partial_{x} u=0$  and the equation (8.3) reads:

\begin{eqnarray}
\frac{1}{\rho}\frac{\partial p} {\partial x} 
=
\frac{\partial }{\partial y}\nu \frac{\partial u}{\partial y}   -\frac{(d+3)(d+5)}{d}\frac{\tau_{T}^{2}{\cal E}}{{\cal K}}\frac{\partial u}{\partial y}\frac{\partial K}{\partial y}
\end{eqnarray}

\noindent The last contribution to the right side of equation (8.6) describing interaction of the energy flux in the wave-number space  and momentum flux in the physical space,  is not small in the proximity to  the walls.   This effect  does not appear in familiar turbulence models. 

\noindent The equations (8.1) - (8.5) are quite involved and,  as in the theories based on renormalized Wyld's expansions of the Navier-Stokes equations,  accounting for the higher-order contributions  to the hydrodynamic approximation is impossible. 
{\it However,  the main difference between  the approach developed in this paper  and the theories   based on the renormalized perturbation expansions  of the Navier-Stokes equations  is that here 
the problem of resummaion of an infinite series does arise:   the equations (8.1)-(8.5) and all high-order 
models,  we even cannot write down,   are contained  in a simple kinetic equation (5.11).  }

\section{Discussions and Conclusions}  

\noindent  The Boltzmann-equation (1.1)-(1.2) contains the Navier-Stokes equations capable of accurate description of turbulent flows. As was demonstrated in the nineties (Yakhot et al 1992), Rubinstein (1990), ) the coarse -graining or small-scale elimination, applied to the NS equations, leads to  the low-order turbulent models. We believe that,  in principle,   {\it the coarse - grained Boltzmann equation (1.1)-(1.2) } contains  coarse-grained hydrodynamic approximations and {\it all}  possible turbulence models (both LES and transport) of an arbitrary nonlinearity and complexity.  At the present time,  this coarse-graining procedure does not exist.  One can attempt to  develop the Wyld diagrammatic expansion applied directly on the non-linear equation (1.1)-(1.2),  eliminate fast modes  and, as a result,  derive turbulence  models. This well-defined  program  is extremely complex and we were unable to achieve much progress.  

Instead, the theory proposed in this  paper is based on a simple qualitative  relaxation time approximation (1.1),(1.3) combined with the `turbulent relaxation time ' $\tau_{T}\approx {\cal K}/{\cal E}$ which is an exact consequence of the energy balance (4.6).  Since kinetic energy is the large-scale property of turbulence,  this relaxation time is the largest time-scale in a flow  leading to the largest `turbulent viscosity' $\nu_{T}=O(v'_{rms}L)=O({\cal K}^{2}/{\cal E}$. This way, the fast modes are overdamped and the resulting equations correspond to transport turbulent modeling describing velocity fluctuations  on the scales  $r\approx L$. 

\noindent {\it 1.  Transport modeling vs LES.}    The  choice of the often - used Maxwell-Gibbs  distribution  function (2.1) is an important step in  derivation of the Navier-Stokes equations from kinetic theory.  Given these  equations,  one defines a   spatial cut- off $\Delta$ and,  averaging over  the small-scale ( $r<\Delta$) fluctuations,  derives the coarse-grained NS equations,  called turbulent models (Yakhot (1992),  Smith  et al (1992), Rubistein et al (1990), Yoshizawa (1987)).   By construction, the resulting equations  for the "resolved scales"  involve  spatial derivatives evaluated on the  mesh (cut -off),  i.e. all gradients in these   equations  are:

\begin{equation}
\frac{\partial u}{\partial x}=\frac{u(x+\Delta/2)-u(x-\Delta/2)}{2\Delta}
\end{equation}

\noindent As was  shown  above,  the  high-order modeling involves high powers  of velocity gradients (velocity differences),  which, due to intermittency ( $\Delta\ll L$),  cannot be expressed in terms of small deviations from the gaussian pdf.   Therefore,  to describe a  flow,  including the inertial range ( $r\approx \Delta \ll L$ ) dynamics,   the model must satisfy an infinite number of non-trivial dynamic constraints  involving, among other terms, pressure-velocity correlations (Yakhot and Sreenivasan (2006)).   With increase of the difference $L-\Delta$,  more and more constraints become relevant and,   at the scales  $r\ll L$,  the quality of all existing LES models  rapidly deteriorate:  with increase of the moment order,  the difference between full  (DNS) and model (LES) simulations   of high-order moments grows. 
 
\noindent  As $\Delta\rightarrow L$,  the probability density of velocity difference is  extremely close to  Gaussian,  the moments  $\overline{(u(x+\Delta/2)-u(x-\Delta/2)^{2n+1}}\rightarrow 0$, $\overline{(u(x+\Delta/2)-u(x-\Delta/2)^{2n}}\rightarrow const$ and  the problem of anomalous scaling does not appear.   (See Fig.). 
This huge simplification was the main and only reason for our choice of the non-equilibrium  probability density  (3.1) leading to the gaussian pdf  (5.7) of the coarse-grained  hydrodynamic field. Thus, the procedure developed here is  valid for   transport modeling only.
To work out  a  similar approach to the LES  ($\Delta\ll L$), one has to use the expressions for the multifractal   pdfs of velocity difference, which is not a simple task.\\

\noindent{\it 2. The  models.}  At the large scales  $r\approx L$,   the zero-order  ($\eta=0$) flow can be described in terms of only {\it two parameters}:  kinetic energy $K$ and mean dissipation rate ${\cal E}$, so that $\tau_{T}\approx K/{\cal E}$. 
Despite superficial similarity of  relaxation times,  the coarse-grained kinetic equation,  derived in this paper,   is very different from the well -known $K-{\cal E}$ model, widely used in engineering:\\
\noindent 1. No  explicitly written  momentum equation is involved in our  model.\\
\noindent 2. No separate equation for kinetic energy $K$  is needed:  all information is contained in the kinetic equation;\\
\noindent 3. The simple equation (5.11) contains  $K-{\cal E}$,  Reynolds stress and all  non-linear models etc. Thus, for example,  it can describe the secondary vortices in a square duct flow, which no $K-{\cal E}$ model is capable of doing.\\
\noindent 4.  No modeling of pressure-velocity correlations is needed:  it is hidden in (5.11). This fact can be of extreme importance for computing complex flows.\\
\noindent 5. The rapid distortion and memory effects  are accounted for  in (5.11).\\
\noindent 6. {\it Sound speed renormalization.} The kinetic equation (5.11)  contains  an equation of state $p'(\rho, \theta)$,  which is a remnant of the integrated out microscopic, close-to- equilibrium,  modes. The equilibrium speed of sound is defined  as usulal: $c_{s}^{2}=(\frac{\partial p'}{\partial \rho})_{S}$.  As follows from (6.12), the speed of sound obtained from  hydrodynamic approximation,   derived in the lowest order of  the non-equilibrium CE expansion,  proposed in this work,  is renormalized:  $\rho c_{s}^{2}=\theta +K$.
This effect, appearing in the higher order of the renormalization group procedure   applied to the NS equation,  was derived and numerically tested in Staroselsky et al (1990) and,
 even  earlier, was   suggested  by Chandrasekhar (1955) who was interested in the role of turbulence in dynamics of  the interstellar gas collapse. 
This feature  may prove important for simulations of turbulent flows. 

\noindent  During last two hundred years,   the  Navier-Stokes equations  enjoyed remarkable success in describing low- Knudsen  and low- Weisenberg number fluid flows. This success was habitually  attributed  to the large scale- separation between hydrodynamic and microscopic modes: $\lambda/L\ll 1$.  There is one caveat in this argument, though.   One must bear in mind   that the NS equations  are {\it closed}  by imposing  proper  equations  of state  relating pressure to density and temperature.   Invariably,  the  equilibrium  equations of state or equilibrium thermodynamic relations are chosen for this purpose. 
The problem is that  the CE procedure,  leading to the NS equations,  is an expansion around the zero-spatial- gradient equilibrium state of fluid and both the pressure gradient  and the gradient of turbulent kinetic energy  are   large -scale properties defined on the same hydrodynamic scales.  Thus,  in this state of fluid, the only relevant dimensionless
 parameter is Mach number $Ma=v_{rms}/c_{s}$.  
Therefore,   to describe the stresses,  in addition to thermodynamic considerations based on an  estimate 
$p=O(\rho c_{s}^{2})$,  one has  to add the turbulent contribution $\delta p\approx \rho v_{rms}^{2}$, which  appears  naturally in the procedure developed above.
Thus,  strictly speaking, the pressure term in the NS equation,  which is the functional of the solution,  must be determined self-consistently.

\noindent 
Given the relaxation time $\tau_{T}\approx K/{\cal E}$ and effective viscosity $\nu_{T}=\tau_{T}K$, the  kinetic equation (5.11) generates  various fields  such as 
${\bf V}$ and $K$. However, to evaluate the relaxation time,  the magnitude of the local 
dissipation rate is needed. 
This can be done selfconsistently,  if, in accord with Kolmogorv theory,  we assume 

\begin{equation}
{\cal E}=\nu_{T}\overline{(\frac{\partial (v-V)_{i}}{\partial x_{j}})^{2}}
\end{equation}

\noindent  Keeping in mind the Lattice Boltzmann applications, we can define the velocity derivatives on a lattice of  lattice constant $\Delta$ with the result written for simplicity in a one-dimensional case:

\begin{equation}
{\cal E}(i)\approx \nu \overline{( \frac{{\bf v}_{i+1}-{\bf v}_{i}+\Delta\partial_{x}{\bf V}(i)}{\Delta})^{2}}
\end{equation}
\noindent where $i$ denotes  position (coordinate)  of a lattice site and the averaging is carried out selfconsistently   on a non-equilibrium pdf,  which is a solution to the kinetic equation (5.11).
Another possibility is to use the dissipation rate directly from the energy balance (8.5). \\

\noindent  The  proportionality  constants  in  (8. 2)    are reasonably close to those quoted in   Speziale (1987),  Rubinstein et al (1990)  and Yoshizawa (1987).  It should be pointed out that the value of  calculations based the low-order trancations is not clear.  Indeed, recasting the results of Sections 7 and  8  in terms of dimensionless rate of strain $\eta\approx K |S_{ij}|/{\cal E}$, we  find that $\sigma^{(2)}\approx \nu^{2} S^{2}/K\approx K\eta^{2}$ and   the next  order gives: $\sigma^{(3)}\propto K\eta^{3}$. Taking into account that in the flows of engineering importance the parameter $\eta$ is not small, often  reaching values $\eta\approx 10-20$, the perturbation expansion    does not converge  and no trancation is, in general,  possible.  {\it We would like to reiterate that the entire series is contained in a relatively simple kinetic equation (5.11)}.

\noindent  The  low-order hydrodynamic approximations and turbulence models, contained in (5.11), are similar to the well-known ones,  which have been widely tested in both scientific and engineering environments (see Chen (2004), for example). 
The future simulations of  strongly nonlinear,   rapidly distorted or oscillating flows,  will provide a decisive test  of the  limits of validity and accuracy of  calculations based on the kinetic equation 
derived in this paper. 

\noindent I am grateful to H. Chen, A. Polyakov, I. Staroselsky, X. Shan, S. Succi,  E. Kamenskaya, T. Gatski, K.R. Sreenivasan   and U. Frisch for most interesting and stimulating discussions.

 \end{document}